\documentclass[12pt]{article}
\usepackage{amsmath,amssymb,amsfonts}
\usepackage{longtable}

\def\mylist{\begin{list}{}{\setlength{\leftmargin}{0.5in}
               \setlength{\listparindent}{-0.5in}
               \setlength{\itemindent}{\listparindent}}}

\newcommand{\bra}[1]{\langle#1\,|}
\newcommand{\ket}[1]{|#1\,\rangle}

\newcommand{\nD}[1]{\not D}
\newcommand{\ZZ}{{\mathbb Z}}

\begin{document}

\begin{titlepage}

\title{BLG theories at low values of Chern-Simons coupling}
\author{Denis Bashkirov\\ {\it\small Perimeter Institute for Theoretical Physics, Waterloo, ON N2L 2Y5, Canada\footnote{E-mail: dbashkirov@perimeterinstitute.ca}}}

\maketitle

\abstract{It was checked in paper \cite{BK} by comparing the moduli spaces and superconformal indices that two of the BLG theories $(SU(2)_{1}\times SU(2)_{-1})/{\mathbb Z}_2$ and $SU(2)_2\times SU(2)_{-2}$ are dual to $U(2)_1\times U(2)_{-1}$ and $U(2)_{2}\times U(2)_{-2}$ ABJM theories, correspondingly. In this paper we consider the BLG theories $SU(2)_1\times SU(2)_{-1}$ and $(SU(2)_2\times SU(2)_{-2})/{\mathbb Z}_2$. These theories were noted in \cite{BK} to be a tensor product of two interacting ${\mathcal N}=8$ SCFT's. In this paper we identify the SCFT's that occur in the product. For both theories one of the sectors is the IR limit of ${\mathcal N}=8$ $SU(2)$ SYM.}

\end{titlepage}

\section{Introduction and summary}

 Several years ago a class of ${\mathcal N}=8$ Superconformal Quantum Field Theories in three dimensions was found by Bagger and Lambert and Gustavsson in \cite{BL,G}. These theories are gauge theories $SU(2)_k\times SU(2)_{-k}$ with gauge group $SU(2)\times SU(2)$ and matter in bifundamental representation. The kinetic term for the gauge field is not the standard Yang-Mills terms but rather the Chern-Simons terms with levels $k$ and $-k$. Soon thereafter Aharony, Bergman, Jafferis and Maldacena found \cite{ABJM} that the low energy dynamics of stacks of $M2$-branes is described by a different class of theories -- $ABJM$ theories which are also Chern-Simons gauge theories with matter, but with gauge groups $U(N)_k\times U(N)_{-k}$ ($N$=2 for the case of two $M2$-branes). Their relation to the low energy dynamics of $M2$-branes is evident from the brane construction, which is absent for $BLG$ theories.

 This motivates an investigation of relations between the two classes of theories. In several papers \cite{ABJ,LP,BK,HH} some relations were proposed. These relations also included $ABJ$  theories \cite{ABJ} that differ from $ABJM$ models in that the ranks of the two $U(N)$ gauge groups are not necessarily the same. In particular, based on the comparison of the moduli spaces and superconformal indices the following identifications were proposed in \cite{BK}

\begin{itemize}
\item $U(2)_1\times U(2)_{-1}$ ABJM theory and $(SU(2)_1\times SU(2)_{-1})/\ZZ_2$ BLG theory
\item $U(2)_2\times U(2)_{-2}$ ABJM theory and $SU(2)_2\times SU(2)_{-2}$ BLG theory
\item $U(3)_2\times U(2)_{-2}$ ABJ theory and $(SU(2)_4\times SU(2)_{-4})/\ZZ_2$ BLG theory
\end{itemize}

It was also noted in the same paper that the $BLG$ theories $SU(2)_1\times SU(2)_{-1}$ and $(SU(2)_2\times SU(2)_{-2})/{\mathbb Z}_2$ are different despite having the same moduli space. Moreover, each of them was noticed to be a reducible theory in the sense that they are tensor products of two SCFT's, although it was not clear what these 'elementary' SCFT's are. In this paper we identify them by computing the superconformal indices. The result is

\begin{itemize}
\item $SU(2)_1\times SU(2)_{-1}$ BLG theory is the product ${\mathcal T}_{SU(2)}\otimes (U(1)_2\times U(1)_{-2})$
\item $(SU(2)_2\times SU(2)_{-2})/{\mathbb Z}_2$ BLG theory is the product ${\mathcal T}_{SU(2)}\otimes{\mathcal T}_{SU(2)}$
\end{itemize}

where $U(1)_2\times U(1)_{-2}$ is the interacting ABJM theory and ${\mathcal T}_{SU(2)}$ is the IR limit of the ${\mathcal N}=8$ Super Yang-Mills with gauge group $SU(2)$.

It was noted in \cite{Distleretal, LT} that moduli spaces of these two BLG theories coincide with the moduli space of the IR limit of maximally supersymmetric $SO(4)$ SYM. From our analysis it follows that BLG theory $(SU(2)_2\times SU(2)_{-2})/{\mathbb Z}_2$ is essentially the IR limit of the maximally supersymmetric $SO(4)\cong(SU(2)\times SU(2))/{\mathbb Z}_2$ SYM, or perhaps of the maximally supersymmetric $Spin(4)\cong SU(2)\times SU(2)$ SYM. The first BLG theory is different.

\section{Existence of two sectors}

 In this section I review and refine the arguments \cite{BK, BK2, B} for the existence of two sectors in BLG theories $SU(2)_1\times SU(2)_{-1}$ and $(SU(2)_2\times SU(2)_{-2})/{\mathbb Z}_2$.

 Given two Quantum Field Theories $QFT_1$ and $QFT_2$ one may construct a third theory $QFT_3\equiv QFT_1\otimes QFT_2$ by taking the tensor product of the two. Thinking about a QFT as a collection of local operators and their OPE's, taking the product just amounts to considering the local operators spanned by products ${\mathcal O}^{(1)}{\mathcal O}^{(2)}(x)$ where the superscripts indicate to which of the two original QFT's the local operator ${\mathcal O}(x)$ belongs. The fact that $QFT_3$ is reducible in this sense is seen, in particular, in the existence of two stress-tensors $T^{(1)}_{\mu\nu}(x)\otimes 1$ and $1\otimes T^{(2)}_{\mu\nu}(x)$. Of course, forming new QFT's in this way is hardly motivated.

 However, it may happen that a seemingly irreducible QFT with some Lagrangian with all fields interacting between themselves can in fact be reducible. One example which will be important for us is the ABJM theory $U(N)_1\times U(N)_{-1}$. It is a Lagrangian theory with superconformal ${\mathcal N}=8$ symmetry in three dimensions. It is not possible to recognize this theory as reducible by just looking at the Lagrangian because the second sector is non-perturbative -- it is realized by monopole operators \cite{BK}. However, it is easy to understand once we know the physical interpretation of this theory -- it describes the low energy dynamics of a stack of $N$ $M2$-branes. As usual, the center of mass of the stack is decoupled, and so the theory is a product of a free theory and an interacting theory. The first factor describes the motion of the center of mass while the second sector describes the interaction between the $N$ of them. It is also known from the relation between $M$-theory and type $IIA$ string theory that this SCFT is the IR limit of the ${\mathcal N}=8$ Super Yang-Mills theory with gauge group $U(N)$. In this theory the $U(1)$ gauge factor decouples, so the center of mass dynamics is described by the IR limit of ${\mathcal N}=8$ $U(1)$ SYM which is known to be the theory of four free chiral scalars in the language of ${\mathcal N}=2$ supersymmetry. The interacting sector is the IR limit of the ${\mathcal N}=8$ $SU(N)$ SYM.

 How do we see the reducibility of the $U(N)_1\times U(N)_{-1}$ ABJM model without invoking its physical interpretation? This was done in \cite{BK2} by analyzing scalar operators in the theory with low conformal dimension. In any ${\mathcal N}=8$ SCFT in three dimensions the stress-tensor supermultiplet has scalar operators with conformal dimension $\Delta=1$ in the representation ${\bf 35}$ of the $R$-symmetry group $SO(8)_R$. The method was to use the deformation proposed by Seok Kim \cite{Kim} of the radially quantized theory (that is, put on the ${\mathbb R}\times {\mathbb S}^2$ appropriately) to get essentially free theory with ${\mathcal N}=2$ superconformal symmetry in backgrounds with magnetic fluxes turned on on ${\mathbb S}^2$, and then in the deformed free theory find the spectrum of chiral (with respect to the unbroken ${\mathcal N}=2$ subgroup of the full superconformal group) scalar operators with conformal dimensions $\Delta=1$ and $\Delta=1/2$. Conformal dimensions of these operators are protected along the deformation, so their spectrum at the two ends of the one-parameter deformation is the same. The protection comes from the independence of the superconformal index on the deformation parameter and the fact the spectrum of the low-$\Delta$ chiral scalars is unambiguously reproduced from the expression for the index. I will give a detailed argument in the next section. The important thing of the analysis was the realization of doubling of chiral scalars spectrum with conformal dimension $\Delta=1$ compared to what one needs for an  ${\mathcal N}=8$ SCFT with one stress-tensor. This meant presence of two stress-tensor multiplets and thus, reducibility of the $U(N)_1\times U(N)_{-1}$ ABJM model. Moreover, presence of scalars with conformal dimension $\Delta=1/2$ implied that one of the sectors was free.

The same analysis was later done in \cite{BK} which showed that both $SU(2)_1\times SU(2)_{-1}$ and $(SU(2)_2\times SU(2)_{-2})/{\mathbb Z}_2$ BLG theories consisted of two sectors both of which are not free but were not identified.

\section{Identification of the two sectors}

In this section we provide the answer to the question of what are the blocks with which the two $SU(2)_1\times SU(2)_{-1}$ and $(SU(2)_2\times SU(2)_{-2})/{\mathbb Z}_2$ BLG theories are built.

The tool that we need is the superconformal index introduced in \cite{BBMR, BM, Kim, IY}. In its simplest form the superconformal index

\begin{equation}
{\mathcal I}(x)=Tr((-1)^Fx^{E+j_3})
\end{equation}
computes the trace taken over the Hilbert space of the theory put on ${\mathbb R}\times{\mathbb S}^2$ with insertions of a number of operators. In fact, only the BPS states for which $\Delta=R+j_3$\footnote{Here $R$ is the generator of the unbroken $U(1)\subset SO(8)_R$.} (we use $\Delta$ and $E$ interchangeably) give nontrivial contribution \cite {BBMR, Kim, IY}. Because of the local operators--states correspondence in Conformal Field Theories the index computes a weighted sum of local operators on ${\mathbb R}^3$. The inserted operators are exponents of extensive quantum numbers, so if a theory $SCFT_3$ is a product of two theories $SCFT_1$ and $SCFT_2$ there must be a factorization of the index

\begin{equation}
{\mathcal I}_3(x)={\mathcal I}_1(x){\mathcal I}_2(x)
\end{equation}

Note that we could insert fugacities corresponding to a number of conserved global charges, but we will not do this to simplify the computations. At the present moment, an explicit expressions for the indices of gauge theories are not known, so we will do Taylor expansion around $x=0$. The resulting expressions will have the form

\begin{equation}
{\mathcal I}(x)=1+A_1x^{1/2}+A_2x+...+A_8x^4+{\mathcal O}(x^{9/2})
\end{equation}

The coefficient $A_1$ counts the number of free chiral (or BPS) scalars with conformal dimension $\Delta=1/2$ and the coefficient $A_2$ gives the number of chiral scalars with conformal dimension $\Delta=1$. This follows from the fact that $j_3$ in the definition of the index is always non-negative, $E$ is always positive except for the vacuum and the unitarity constraints. Indeed, for BPS operators which are the ones that give nontrivial contributions to the index the matrix element $\bra{BPS}\{Q,S\}\ket{BPS}=\bra{BPS}(E-R-j_3)\ket{BPS}=0$ is zero, where $Q$ is one of the supercharges and $S$ is one of the superconformal charges which is conjugate to $Q$ in the radial quantization. If $j_3$ were negative, then choosing the supercharge $Q'$ with spin opposite to that of $Q$ we would get $\bra{BPS}\{Q',S'\}\ket{BPS}=\bra{BPS}(E-R+j_3)\ket{BPS}<0$ which cannot be in a unitary theory because $\bra{BPS}\{Q',S'\}\ket{BPS}=\bra{BPS}(Q'(Q')^\dagger+(Q')^\dagger Q')\ket{BPS}$ must be nonnegative.

Because the BLG theories we consider do not contain free sectors, the coefficient $A_1$ will be zero. The coefficient $A_2$ is $10$ for an irreducible ${\mathcal N}=8$ SCFT theory and $20$ if the theory consists of two ${\mathcal N}=8$ SCF sectors \cite{BK}.

We start with the computation of the superconformal index for the IR limit of the ${\mathcal N}=8$ $SU(2)$ super Yang-Mills theory. This cannot be done directly as the formula for the
index worked out in \cite{Kim, IY} require the IR R-symmetry to be the UV R-symmetry.\footnote{No mixing with global symmetries is possible as the R-symmetry group is nonabelian $SO(8)_R$.}
Nevertheless, we can employ a round-about tactics. ${\mathcal N}=8$ $U(2)$ super Yang-Mills is dual in IR to the $U(2)_1\times U(2)_{-1}$ model whose index can be computed and gives the result

\begin{align}
{\mathcal I}_{ABJM}(x)=1+4x^{1/2}+20x+56x^{3/2}+139x^2+260x^{5/2}+436x^3+640x^{7/2}+954x^4+{\mathcal O}(x^{9/2})
\end{align}
See the appendix for the details of all the calculations.

As we argued above ${\mathcal I}_{ABJM}={\mathcal I}_{U(2)}(x)={\mathcal I}_{U(1)}(x){\mathcal I}_{SU(2)}(x)$. The $U(1)$ ${\mathcal N}=8$ super Yang-Mills in the IR is equivalent to four free ${\mathcal N}=2$ chiral multiplets whose index is easy to compute

\begin{align}
{\mathcal I}_{4chirals}(x)=1+4x^{1/2}+10x+16x^{3/2}+19x^2+20x^{5/2}+26x^3+40x^{7/2}+49x^4+{\mathcal O}(x^{9/2})
\end{align}

Dividing one by the other we get the superconformal index for the IR limit of ${\mathcal N}=8$ $SU(2)$ SYM

\begin{align}
{\mathcal I}_{SU(2)}(x)=1+10x+20x^2+20x^3+65x^4+{\mathcal O}(x^{9/2})
\end{align}

The superconformal index for the $(SU(2)_2\times SU(2)_{-2})/{\mathbb Z}_2$ BLG theory is

\begin{align}
{\mathcal I}_{BLG_2}(x)=1+20x+140x^2+440x^3+930x^4+{\mathcal O}(x^{9/2})
\end{align}

This amounts to

\begin{align}
{\mathcal I}_{BLG_2}(x)={\mathcal I}_{SU(2)}^2(x)
\end{align}
 up to order $x^4$.

This gives rise to the conjecture

\begin{itemize}
\item $(SU(2)_2\times SU(2)_{-2})/{\mathbb Z}_2$ BLG theory is equivalent to two copies of the IR limit of ${\mathcal N}=8$ $SU(2)$ Super Yang-Mills.
\end{itemize}

To propose the second equivalence we first need to compute the index of the $U(1)_2\times U(1)_{-2}$ ABJM model. The result is

\begin{align}
{\mathcal I}_{ABJM'}(x)=1+10x+19x^2+26x^3+49x^4+{\mathcal O}(x^{9/2})
\end{align}

The index for $SU(2)_1\times SU(2)_{-1}$ BLG theory is

\begin{align}
{\mathcal I}_{BLG_1}(x)=1+20x+139x^2+436x^3+954x^4+{\mathcal O}(x^{9/2})
\end{align}

The multiplication gives

\begin{align}
{\mathcal I}_{BLG_1}={\mathcal I}_{ABJM'}(x){\mathcal I}_{SU(2)}(x)
\end{align}
up to the fourth order in $x$.

This leads to the second conjecture

\begin{itemize}
\item $SU(2)_1\times SU(2)_{-1}$ BLG theory is equivalent to a product of two theories: $U(1)_2\times U(1)_{-2}$ ABJM model and the IR limit of ${\mathcal N}=8$ $SU(2)$ Super Yang-Mills.
\end{itemize}

\section{Moduli spaces}

According to the conjecture from the previous section the moduli spaces in both cases should be
\begin{equation}
{\mathbb C}^4/{\mathbb Z}_2\times {\mathbb C}^4/{\mathbb Z}_2
\end{equation}
which is the product of two moduli spaces ${\mathbb C}^4/{\mathbb Z}_2$. Each ${\mathbb Z}_2$ changes the sign of the vectors in the corresponding copy of ${\mathbb C}^4$.

Note that the analysis of \cite{LT, Distleretal, LP} yielded the following moduli space for $SU(2)_1\times SU(2)_{-1}$ and $(SU(2)_2\times SU(2)_{-2})/{\mathbb Z}_2$
\begin{equation}
\frac{{\mathbb C}^4\times{\mathbb C}^4}{{\mathbb Z}_2\times{\mathbb Z}_2}
\end{equation}

where one of ${\mathbb Z}_2$ swaps the two copies of ${\mathbb C}^4$ and the other changes the signs of elements of both copies. If we denote the coordinates in the two copies of ${\mathbb C}^4$ az $(w_1,w_2)$, then in coordinates $(z_1=w_1+w_2,z_2=w_1-w_2)$ the action of the group ${\mathbb Z}_2\times{\mathbb Z}_2$ is generated by operations $(z_1,z_2)\to(z_1,-z_2)$ and $(z_1,z_2)\to(-z_1,-z_2)$. The same group is generated by two operations $(z_1,z_2)\to(z_1,-z_2)$ and $(z_1,z_2)\to(-z_1,z_2)$ which correspond to the moduli space ${\mathbb C}^4/{\mathbb Z}_2\times {\mathbb C}^4/{\mathbb Z}_2$. So the moduli spaces coincide.

\newpage

\section*{Appendix}

The formula that computes the superconformal index was obtained using the localization technique in \cite{Kim, IY}. It has the following form

\begin{align}
{\cal I}(x)=\sum_{\{n_i\}}\int[da]_{\{n_i\}}x^{E_0(n_i)}e^{S_{CS}^0(n_i,a_i)}exp(\sum_{m=1}^\infty f(x^m,ma_i))
\end{align}
 The sum runs over GNO charges \cite{GNO}, the integral whose measure depends on GNO charges is over a maximal torus of the gauge group, $E_0(n_i)$ is the energy of a bare monopole with GNO charges $\{n_i\}$, $S_{CS}^0(n_i)$ is effectively the weight of the bare monopole with respect to the gauge group and the function $f$ depends on the content of vector multiplets and hypermultiplets. For details see \cite{Kim, IY}.

All indices are computed up to the fourth order in $x$.

\begin{longtable}{|l|l|}
\hline
Topological charge & Index contribution\\
\hline
$T=0$ & $1+4x+x^2+4x^3+7x^4$\\
\hline
$T=1$& $3x+4x^2+8x^4$\\
\hline
$T=2$& $5x^2+4x^3$\\
\hline
$T=3$& $7x^3+4x^4$\\
\hline
$T=4$ & $9x^4$\\
\hline
total & $1+10x+19x^2+26x^3+49x^4$\\
\hline
\caption{$U(1)_2\times U(1)_{-2}$}
\end{longtable}

\begin{longtable}{|l|l||l|l|}

  \hline
GNO charges & Index contribution & GNO charges & Index contribution\\
  \hline
   $T=0$ & & $T=5$ & \\
 \hline
 $\ket{0,0}\ket{0,0}$ & $1+4x+12x^2+8x^3+12x^4$ & $\ket{3,2}\ket{3,2}$ & $2(6x^{\frac52}+14x^{\frac72})$ \\
 $\ket{1,-1}\ket{1,-1}$ & $4x+16x^2+16x^3+33x^4$ & $\ket{4,1}\ket{4,1}$ & $2(5x^{\frac52}+14x^{\frac72})$\\
 $\ket{1,-1}\ket{0,0}$ & $-x^4$ & $\ket{5,0}\ket{5,0}$ & $2(3x^{\frac52}+14x^{\frac7/2})$\\
 $\ket{0,0}\ket{1,-1}$ & $-x^4$ & $\ket{6,-1}\ket{6,-1}$ & $14x^{\frac72}$\\
 $\ket{2,-2}\ket{2,-2}$ & $9x^2+24x^3+16x^4$ & &\\
 $\ket{3,-3}\ket{3,-3}$ & $16x^3+32x^4$ & &\\
 $\ket{4,-4}\ket{4,-4}$ & $25x^4$ & &\\
  \hline
  $T=1$ & & $T=6$ & \\
 \hline
 $\ket{1,0}\ket{1,0}$ & $2(x^{\frac12}+6x^{\frac32}+10x^{\frac52}+7x^{\frac72})$ & $\ket{3,3}\ket{3,3}$ & $10x^3+16x^4$\\
 $\ket{2,-1}\ket{2,-1}$ & $2(3x^{\frac32}+10x^{\frac52}+8x^{\frac72})$ & $\ket{4,2}\ket{4,2}$ & $15x^3+32x^4$\\
 $\ket{3,-2}\ket{3,-2}$ & $2(6x^{\frac52}+14x^{\frac72})$ & $\ket{5,1}\ket{5,1}$ & $12x^3+32x^4$\\
 $\ket{4,-3}\ket{4,-3}$ & $20x^{\frac72}$ & $\ket{6,0}\ket{6,0}$ & $7x^3+32x^4$\\
 & &  $\ket{7,-1}\ket{7,-1}$ & $16x^4$\\
  \hline
$T=2$ & & $T=7$ &\\
\hline
$\ket{1,1}\ket{1,1}$ & $3x+8x^2+12x^3+8x^4$ & $\ket{4,3}\ket{4,3}$ & $20x^{\frac72}$\\
$\ket{2,0}\ket{2,0}$ & $3x+16x^2+19x^3+24x^4$ & $\ket{5,2}\ket{5,2}$ & $18x^{\frac72}$\\
$\ket{3,-1}\ket{3,-1}$ & $8x^2+24x^3+16x^4$ & $\ket{6,1}\ket{6,1}$ & $14x^{\frac72}$\\
$\ket{4,-2}\ket{4,-2}$ & $15x^3+32x^4$ & $\ket{7,0}\ket{7,0}$ & $8x^{\frac72}$\\
$\ket{5,-3}\ket{5,-3}$ & $24x^4$ & &\\
\hline
$T=3$ & & $T=8$ & \\
\hline
$\ket{2,1}\ket{2,1}$ & $2(3x^{\frac32}+10x^{\frac52}+9x^{\frac72})$ & $\ket{4,4}\ket{4,4}$ & $15x^4$\\
$\ket{3,0}\ket{3,0}$ & $2(2x^{\frac32}+10x^{\frac52}+10x^{\frac72})$ & $\ket{5,3}\ket{5,3}$ & $24x^4$\\
$\ket{4,-1}\ket{4,-1}$ & $2(5x^{\frac52}+14x^{\frac72})$ & $\ket{6,2}\ket{6,2}$ & $21x^4$\\
$\ket{5,-2}\ket{5,-2}$ & $18x^{\frac72}$ & $\ket{7,1}\ket{7,1}$ & $16x^4$\\
 & & $\ket{8,0}\ket{8,0}$ & $9x^4$\\
\hline
$T=4$ & \\
\hline
$\ket{2,2}\ket{2,2}$ & $6x^2+12x^3+12x^4$ & &\\
$\ket{3,1}\ket{3,1}$ & $8x^2+24x^3+16x^4$ & &\\
$\ket{4,0}\ket{4,0}$ & $5x^2+24x^3+21x^4$ & &\\
$\ket{5,-1}\ket{5,-1}$ & $12x^3+32x^4$ & &\\
$\ket{6,-2}\ket{6,-2}$ & $21x^4$ & &\\
\hline
\caption{$U(2)_1\times U(2)_{-1}$.
$T$ stands for the topological charge.}
\end{longtable}

\begin{longtable}{|l|l|}
  \hline
GNO charges & Index contribution\\
  \hline
$\ket{0}\ket{0}$ & $1+4x+40x^2+76x^3+114x^4$\\
$\ket{1}\ket{1}$ & $10x+64x^2+132x^3+225x^4$\\
$\ket{2}\ket{2}$ & $35x^2+144x^3+196x^4$\\
$\ket{3}\ket{3}$ & $4(21x^3+64x^4)$\\
$\ket{4}\ket{4}$ & $165x^4$\\
$\ket{1}\ket{0}$ & $-x^4$\\
$\ket{0}\ket{1}$ & $-x^4$\\
\hline
\caption{$SU(2)_1\times SU(2)_{-1}$}
\end{longtable}

\begin{longtable}{|l|l|}
  \hline
GNO charges & Index contribution\\
  \hline
$\ket{0}\ket{0}$ & $1+10x+40x^2+76x^3+114x^4$\\
$\ket{1/2}\ket{1/2}$ & $10x+65x^2+136x^3+199x^4$\\
$\ket{1}\ket{1}$ & $35x^2+144x^3+196x^4$\\
$\ket{3/2}\ket{3/2}$ & $4(21x^3+64x^4)$\\
$\ket{2}\ket{2}$ & $165x^4$\\
$\ket{1}\ket{0}$ & $0$\\
$\ket{0}\ket{1}$ & $0$\\
\hline
\caption{$(SU(2)_2\times SU(2)_{-2})/\ZZ_2$}
\end{longtable}

\section{Acknowledgements}

I thank Anton Kapustin for useful correspondence. This work is supported by the Perimeter Institute for Theoretical Physics. Research at the Perimeter Institute is supported by the Government of Canada through Industry Canada and by Province of Ontario through the Ministry of Economic Development and Innovation.

\end{document}